\documentclass[a4paper,11pt]{article}
\usepackage{aaskaiid}
\usepackage{orcidlink}
\usepackage{hyperref}
\usepackage{cleveref}

\usepackage{xspace}

\newcommand{\review}[1]{\textcolor{black}{#1}}

\title{Probing the Baryon Distribution with Fast Radio Bursts}
\ShortTitle{FRBs: Tracing the Baryon Distribution}

\ShortName{Caleb et al.} 
\author[1]{Manisha Caleb\orcidlink{0000-0002-4079-4648}}
\author[2,3]{Jéferson A. S. Fortunato\orcidlink{0000-0001-7983-1891}}
\author[4,5]{Steffen Hagstotz\orcidlink{0000-0002-7044-3793}}  
\author[6]{Clancy W. James\orcidlink{0000-0002-6437-6176}}
\author[7,8]{Joscha N. Jahns-Schindler\orcidlink{0000-0003-4193-6158}}
\author[9]{Dylan L. Jow\orcidlink{0000-0003-3236-8769}}
\author[10]{Evan F. Keane\orcidlink{0000-0002-4553-655X}}
\author[11]{Koustav Konar\orcidlink{0000-0002-8236-1605}}
\author[11]{Yin-Zhe Ma\orcidlink{0000-0001-8108-0986}}
\author[12]{Daniele Michilli\orcidlink{0000-0002-2551-7554}}
\author[13]{Robert Reischke$^*$\orcidlink{0000-0001-5404-8753}}
\author[14]{Amit Seta\orcidlink{0000-0001-9708-0286}}
\author[15]{Priyanka Singh$^*$}
\author[16]{Laura G. Spitler$^*$\orcidlink{0000-0002-3775-8291}}
\author[17]{Yidan Wang\orcidlink{0000-0002-7372-4160}} 
\author[2,3]{Amanda Weltman\orcidlink{0000-0002-5974-4114}}
\author[]{The SKA Transients SWG}

\affiliation[1]{Sydney Institute for Astronomy, School of Physics, The University of Sydney, Sydney, NSW 2006, Australia}
\affiliation[2]{High Energy Physics, Cosmology \& Astrophysics Theory (HEPCAT) Group, Department of Mathematics and Applied Mathematics, University of Cape Town, Cape Town, 7700, South Africa}
\affiliation[3]{African Institute for Mathematical Sciences, 6 Melrose Road, Muizenberg, Cape Town, 7945, South Africa}
\affiliation[4]{Universit\"ats-Sternwarte, Fakult\"at f\"ur Physik, Ludwig-Maximilians Universit\"at M\"unchen, Scheinerstraße 1, D-81679 M\"unchen, Germany and}
\affiliation[5]{
Excellence Cluster ORIGINS, Boltzmannstraße 2, D-85748 Garching, Germany}
\affiliation[6]{International Centre for Radio Astronomy Research, Curtin University, Bentley, 6102, WA, Australia}
\affiliation[7]{Centre for Astrophysics and Supercomputing, Swinburne University of Technology, Hawthorn, VIC 3122, Australia}
\affiliation[8]{ARC Centre of Excellence for Gravitational Wave Discovery (OzGrav), Hawthorn, VIC 3122, Australia}
\affiliation[9]{Kavli Institute for Particle Astrophysics and Cosmology, Stanford University, 452 Lomita Mall, Stanford, CA 94305, USA}
\affiliation[10]{School of Physics, Trinity College Dublin, College Green, Dublin 2, D02 PN40, Ireland}
\affiliation[11]{Department of Physics, Stellenbosch University, Matieland 7602, South Africa}
\affiliation[12]{Laboratoire d’Astrophysique de Marseille, Aix-Marseille University, CNRS, CNES, Marseille, France}
\affiliation[13]{Argelander-Institut f\"ur Astronomie, Universit\"at Bonn, Auf dem H\"ugel 71, D-53121 Bonn, Germany}
\affiliation[14]{Research School of Astronomy and Astrophysics, Australian National University, Canberra, ACT 2611, Australia}
\affiliation[15]{Department of Astronomy, Astrophysics and Space Engineering, Indian Institute of Technology, Indore 453552, India}
\affiliation[16]{Max-Planck-Institut f\"ur Radioastronomie, Auf dem H\"ugel 69, 53121 Bonn, Germany}
\affiliation[17]{National Astronomical Observatories, Chinese Academy of Sciences, Beijing 100101, China University of Chinese Academy of Sciences, Beijing 100049, China}
\affiliation[*]{Coordinating author}
\emailAdd{reischke@posteo.net}
\emailAdd{lspitler@mpifr-bonn.mpg.de}

\abstract{Baryonic feedback redistributes matter on small to mid cosmological scales, ultimately limiting inferences from Stage IV galaxy surveys. Direct baryon tracers are crucial for recovering cosmological signals masked by astrophysical effects, and vice versa: galaxy formation and other astrophysical processes must be interpreted cosmologically. 
Fast radio bursts (FRBs) serve as such tracers: their dispersion measure (DM) records the line-of-sight integrated ionised electron density. The Square Kilometre Array (SKA) will be the only radio telescope capable of detecting many FRBs in the southern hemisphere, significantly enhancing synergy with surveys such as Rubin Observatory. This chapter completes the FRB trilogy by forecasting the SKA's potential to constrain the baryon distribution from cosmological to galactic scales and across cosmic time.

We tackle this question by investigating the DM scatter as a function of redshift. We also study the statistical properties of the DM field and its cross-correlation with Stage IV galaxy surveys. Our focus is on cosmic shear and galaxy clustering. This shows that the SKA can play a crucial role in pinpointing baryonic feedback models, thereby greatly enhancing the cosmological constraining power of Stage IV galaxy surveys. Furthermore, we show that the SKA will be able to measure the properties of the circumgalactic medium using the scattering timescale of FRBs. Lastly, the large redshift range of FRB detections with the SKA can improve our understanding of the epoch of reionisation. It may also clarify the mechanism behind FRBs.}

\begin{document}
\maketitle

\section{Introduction}
Fast radio bursts (FRBs) are the first brief radio signals from distant parts of the universe that last only milliseconds. Their high energy allows them to be detected even at large distances. 
The bursts interact with free electrons along the line of sight. This leads to dispersion that depends on the number of interactions and, hence, on the integrated electron density, a quantity called the dispersion measure (DM). FRBs are a unique cosmological probe in several respects. First, they are the only truly direct probe of the (ionised) visible matter in the Universe. Second, they can be studied at the cosmological background level, probing the mean baryon density and the expansion of the Universe. Additionally, one can also study perturbations of the DM, similar to weak gravitational lensing.

Since FRBs can be detected in any direction, the Square Kilometre Array (SKA) will primarily conduct commensal observations alongside other surveys. 
Compared with current survey instruments 
\citep[e.g. ASKAP, CHIME, DSA‑110, MeerKAT][]{Hotan2021,Shannon2024,CHIME2021,CHIME2022,Law2023,Rajwade+22,caleb_2023_subarcsec,2025arXiv250801648C}, 
SKA‑Mid will offer substantially greater sensitivity, albeit with a smaller field of view than some surveys. 
This yields detection numbers comparable to those of CHIME, but with a higher mean redshift. Furthermore, SKA‑Low ($50-350\,$MHz) probes an underexplored frequency band, making predictions uncertain. SKA’s sensitivity makes it especially powerful for discovering high‑redshift FRBs. These are of particular interest for cosmological applications due to the cumulative effect of the DM.

The SKA Science Book includes two companion chapters on FRBs. In \citet{Curtin01.2026.SKA}, SKA's possibilities of detecting the central engine(s) behind FRBs are discussed in more detail. \citet{Caleb02.2026.SKA} focuses on the background evolution of the Universe, tests of fundamental physics, and observables beyond the DM. In contrast, this chapter investigates what FRBs detected by the SKA can reveal about the distribution of baryons in the Universe, as well as the related observational requirements.
For details on the number of FRBs detectable with the SKA and the basics of FRB observables, we refer to \citet{Caleb02.2026.SKA}. Specifically, we will investigate how the statistical properties of the baryon, and therefore the electron distribution, are inherited by the DM. We will then explore how this can be used to unlock new constraints on their distribution in the Universe, possibly up to reionisation, using the DM, scattering, and their cross-correlation with other surveys.

This chapter is organised as follows: in \Cref{sec:general_discussion}, we will briefly recap how FRBs measure the baryon content in the Universe. \Cref{sec:macquart_relation_feedback} illustrates how feedback can be constrained from the uncertainty model of the DM redshift relation. SKA's potential to study feedback in synergy with optical surveys is explored in \Cref{sec:two_point_feedback}. We close by looking into the scattering by the cool gas in the CGM and the epoch of reionisation in \Cref{sec:CGMscattering,sec:eor}.

\section{Distribution of ionised baryons from galactic to cosmological scales}
\label{sec:general_discussion}
\review{Due to its collisionless nature, the distribution of dark matter alone in the Universe is reasonably well understood in both the linear and non-linear regimes, forming the hierarchical cosmic web with dark matter halos in collapsed overdense regions.
The distribution of baryons, on the other hand, remains uncertain mainly due to the highly complex non-gravitational interactions. While baryons make up only a small fraction of the total matter, they influence key processes relevant to both galaxy formation and cosmology. Even more so, redistributing the baryons also displaces dark matter via gravitational back-reaction. Therefore, these complex interactions reshape the distribution of baryons and dark matter and therefore have a large influence on the total matter distribution in the Universe.
Observational probes that directly trace the baryon density are therefore key for understanding this visible part of the Universe. 
}
\review{
The ability to investigate cosmological structures on non-linear scales is thus fundamentally limited by uncertainties associated with the complex interactions described above. These are often referred to as baryonic feedback, which changes the statistical properties of large-scale structure (LSS) in the non-linear regime \citep[see][for works on simulations and overviews]{Rudd_pkSuppression_2008, 2011MNRAS.417.2020S, VanDaalen2011, Chisari_Pk2019, VanDaalen_Pk2020, Schaye2023_flamingo}.} Crucially, the precise type of interactions and the extent to which the distribution of baryons is altered in the LSS depend on physical processes that are not resolved by the simulations, known as subgrid physics. 
These phenomena hence pose substantial modelling challenges for measurements carried out by Stage IV galaxy surveys such as Euclid\footnote{\url{https://www.esa.int/Science_Exploration/Space_Science/Euclid}}, the Rubin Observatory Legacy Survey of Space and Time (LSST)\footnote{\url{https://www.lsst.org/}}, and the Roman Space Telescope\footnote{\url{https://roman.gsfc.nasa.gov/}}, which rely on a precise understanding of matter clustering across a wide range of scales. In particular, these surveys obtain most of their signal from scales affected by feedback processes.
Hence, any external measurement of the baryon distribution is paramount. The most common probes are the kinetic (kSZ) and thermal (tSZ) Sunyaev-Zel’dovich (SZ) effect, which have already been used in \citet{Schaan_velRec_2016,Hand:2012ui,Planck:2015ywj,DES:2016umt,Schaan2021_sz, Amodeo2021_sz,2022A&A...660A..27T,2025arXiv250707991K} to infer the distribution of baryons statistically. Furthermore, $X$-ray data are increasingly competitive at discriminating between different feedback models \citep{Schneider2022_WL_xray_kSZ, Bigwood2024, Grandis2023, 2024arXiv240707152H, McCarthy_kSZ2024,2025arXiv251202954S}.
Every cosmological probe has its systematic factors. The above are either limited by their sensitivity to gas temperature (e.g., tSZ, X-ray) or heavily depend on the assumed metallicity distribution and ionisation mechanism (e.g., absorption line studies). New probes that can overcome these drawbacks and have different systematic effects are therefore essential. 

The observed DM is given by a number of contributions: $\mathrm{DM}_{\mathrm{host}}$, the host contribution originating from the burst's host galaxy as well as from its local environment, $\mathrm{DM}_{\mathrm{cosmic}}$, the contribution from the mean free electron density and the LSS in the Universe, $\mathrm{DM}_{\mathrm{MW}}(\mathbf{x})$ and $ \mathrm{DM}_{\mathrm{MW,halo}}(\mathbf{x})$ are the contributions from the Milky Way and its halo:
\begin{equation}
\label{eq:parts}
    \mathrm{DM}_\mathrm{obs}(\mathbf{x},z) = \mathrm{DM}_{\mathrm{host}}(z) + \mathrm{DM}_{\mathrm{cosmic}} (z,\mathbf{x})+ \mathrm{DM}_{\mathrm{MW}}(\mathbf{x}) + \mathrm{DM}_{\mathrm{MW,halo}}(\mathbf{x})\,.
\end{equation}
The cosmological contribution to the DM for an FRB at redshift $z$ and in direction $\hat{\boldsymbol{x}}$ can be written as follows:
\begin{equation}
\label{eq:DM_LSS_v2}
    \mathrm{DM}^{\mathstrut}_\mathrm{cosmic}(\boldsymbol{x},z) = \frac{3 \Omega_\mathrm{b0} \chi_\mathrm{H}}{8 \uppi G m_\mathrm{p}}\chi_\mathrm{e}  \int_0^z   \frac{1+z'}{E(z')} f^{\mathstrut} (z^\prime) \big[1+\delta^{\mathstrut}_\mathrm{e}(\boldsymbol{x},z')\big] \mathrm{d} z' ,
\end{equation}
with the dimensionless baryon density parameter $\Omega_\mathrm{b0}$ today, the Hubble radius, $\chi_\mathrm{H} = c/H_0$, the proton mass $m_\mathrm{p}$, the number of electrons per baryon $\chi_\mathrm{e}$ and the gravitational constant $G$. $\delta_\mathrm{e}$ is the comoving electron density contrast, $E(z)$ is the expansion function, and $f(z)$ is the fraction of ionised baryons.
\Cref{eq:DM_LSS_v2} can be studied in two ways: the mean relation\footnote{Called the Macquart relation in memory of the late and great J.P. Macquart.} 
\begin{equation}
\label{eq:macquart_relation}
    \langle\mathrm{DM}^{\mathstrut}_\mathrm{cosmic}\rangle (z) = \frac{3 \Omega_\mathrm{b0} \chi_\mathrm{H}}{8 \uppi G m_\mathrm{p}}\chi_\mathrm{e}  \int_0^z   \frac{1+z'}{E(z')} f^{\mathstrut} (z^\prime)\mathrm{d}z^\prime\;,
\end{equation}
contains information about the expansion history of the Universe, as well as its mean baryon density. This has been studied in many papers, e.g. \citet{macquart_census_2020,2022MNRAS.511..662H,2022MNRAS.516.4862J} to name a few prominent ones. The fluctuations of the DM, which are sourced by $\delta_\mathrm{e}$, create the uncertainty on the Macquart relation and already contain information about feedback at this level \citep[see e.g.][]{2025arXiv250717742R,2025ApJ...989...81S}. However, this approach does not distinguish among different spatial scales, treating uncertainty as a point estimate. 
If, however, we aim to recover the spatial distribution of baryons in the Universe, summary statistics that can quantify the amount of correlation in the DM as a function of spatial scale should be studied. In the following, we will discuss both cases in \Cref{sec:macquart_relation_feedback} and \Cref{sec:two_point_feedback} respectively.

\section{The distribution of baryons from the Macquart relation}
\label{sec:macquart_relation_feedback}

\begin{figure}
    \centering
    \includegraphics[width=0.75\linewidth]{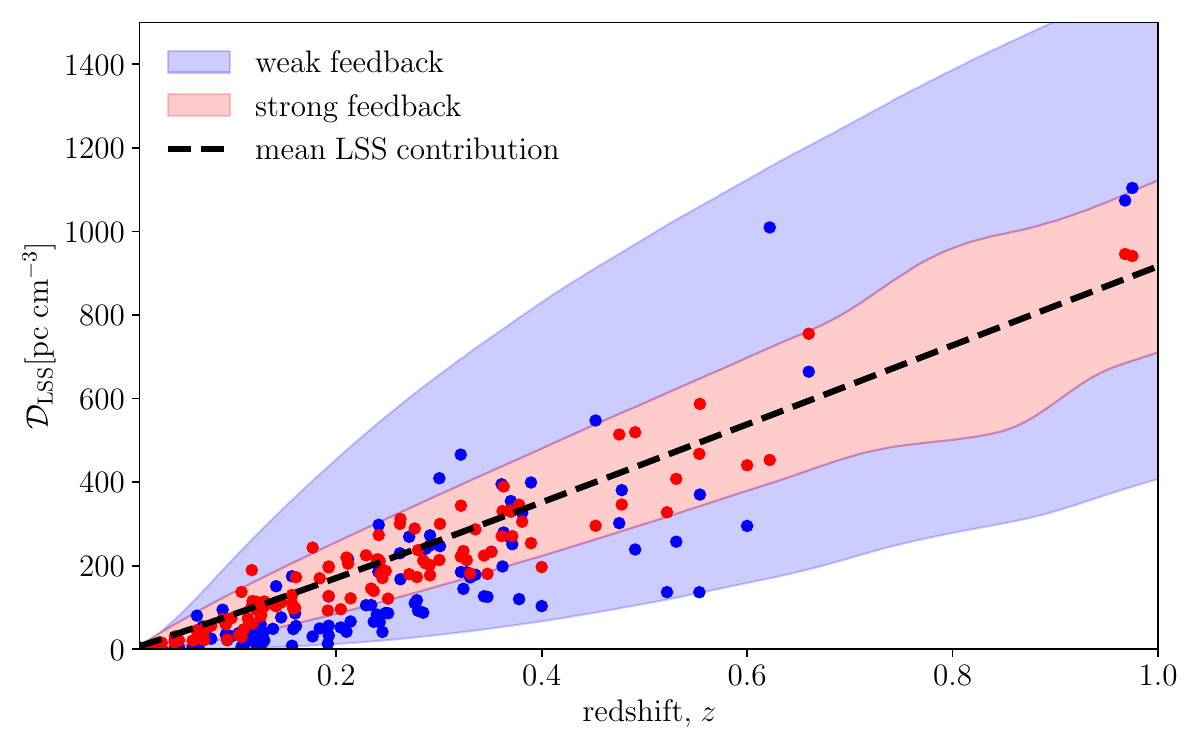}
    \caption{\review{Mock observations of the Macquart relation for 100 FRBs with varying feedback strength for illustration. Stronger baryonic feedback (red) redistributes baryons and leads to a smoother electron distribution, leading to decreased scatter in the Macquart relation. For clarity, we also show the contour level where the corresponding likelihood has dropped to 5\% of its peak value.}}
    \label{fig:pdf_feedback_macquart}
\end{figure}

\Cref{eq:macquart_relation} itself depends only on the mean baryon density. However, the observed DM of two FRBs at the same redshift but in different directions will vary due to cosmic variance and uncertainties in the other contributions. Baryonic feedback in the form of supernovae and AGN activity drives, leading to a smoother distribution of cosmic electrons. \review{The effect on $\mathrm{DM}_\mathrm{cosmic}$ is shown in \Cref{fig:pdf_feedback_macquart}, where stronger feedback leads to a smaller scatter in the Macquart relation as indicated by the data points and the contour of the corresponding likelihoods. 
It can be seen that the likelihood spreads more towards large DM values due to the distribution's long tail, making the distribution more asymmetric. For DM below the mean, the increase in scatter is more gradual but still visible. In general, the changes shown here are driven by variations in the underlying gas profiles that comprise the halos. If gas profiles become shallower, the sightline-to-sightline becomes smaller.
For a more complete characterisation of the interplay of the profiles with the shape of the probability distribution function, we refer to \citet{2026arXiv260118784T}.}

\begin{figure}
    \centering
    \includegraphics[width=0.75\linewidth]{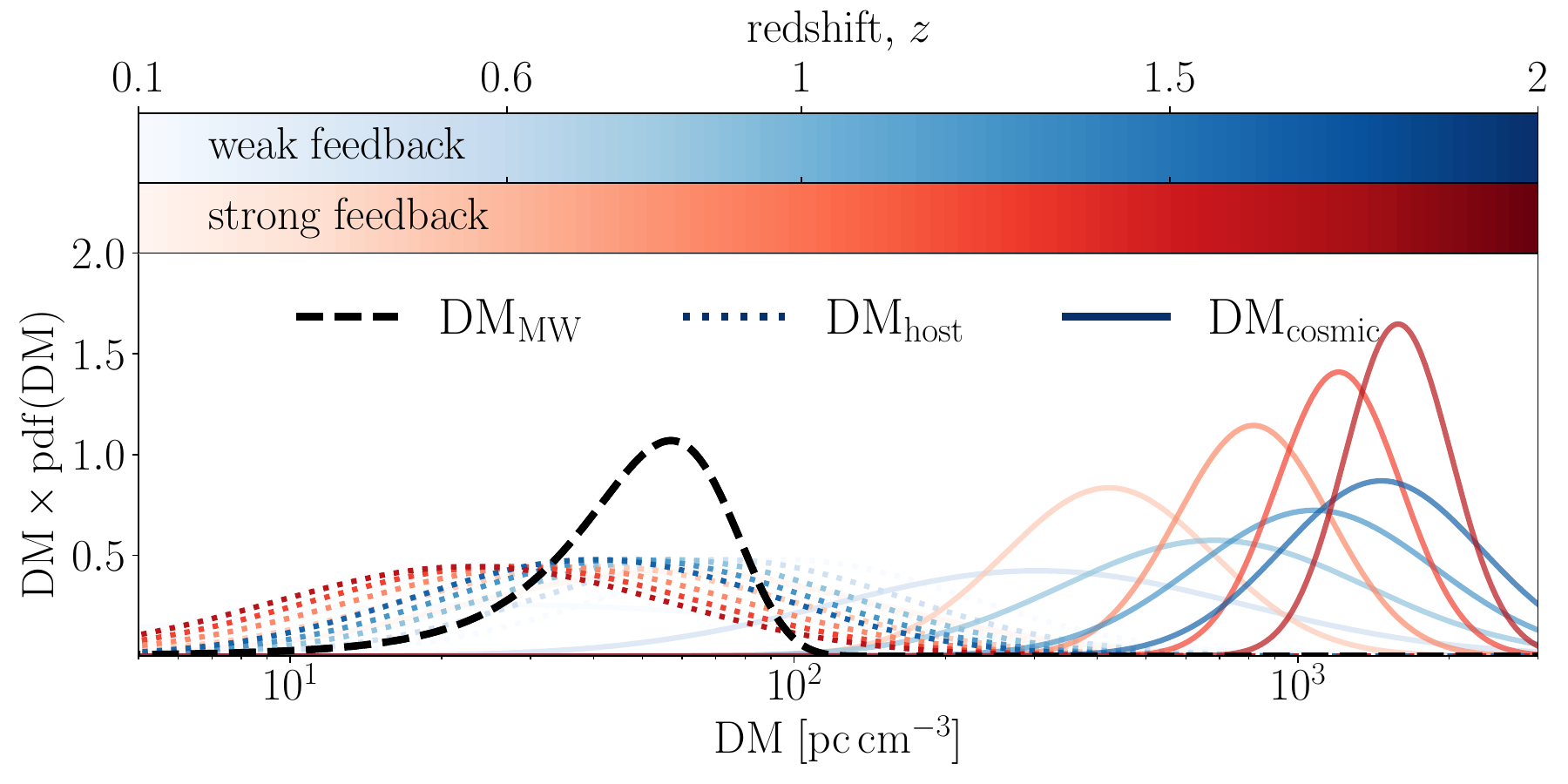}
    \caption{Shown is the scaling of the different components (their probability density function in different line-styles) with redshift from \Cref{eq:parts} depicted as a colour gradient. Red-coloured lines show a scenario with strong feedback (a fairly smooth electron distribution), and blue-coloured lines show one with weak feedback (a very clustered electron distribution).}
    \label{fig:pdf_feedback}
\end{figure}

The detailed evolution of the individual contributions that make up the total DM is illustrated in \Cref{fig:pdf_feedback}, which plots the probability distribution functions (or likelihoods) of the components in \Cref{eq:parts} as a function of redshift for two feedback strengths. The Milky Way contribution, shown as a black dashed line, assumes values between $10^1$ and $10^2$ in DM. However, the exact shape of this curve is rather unknown, and it is usually marginalised over in cosmological analyses of FRBs. The dotted lines indicate the host contribution, which diminishes with increasing redshift. This evolution is mainly due to cosmological redshift rather than an intrinsic evolution of the DM of the host galaxy \citep{2024arXiv240308611T}. As shown analytically in \citet{2025OJAp....8E.127R}, stronger feedback reduces the $\mathrm{DM}_\mathrm{host}$ as the host galaxy will retain less gas, and it is distributed more into the ambient LSS, i.e. the intergalactic medium (IGM). The solid lines show the cosmological contribution, $\mathrm{DM}_\mathrm{cosmic}$; the mean is, by definition, unchanged across different feedback scenarios. However, as feedback increases, the gas distribution becomes smoother, reducing the likelihood's width. This effect was used in \citet{2025arXiv250717742R} to measure baryonic feedback and the power spectrum suppression directly from FRBs. For forecasts of this method, we also refer to \citet{2025ApJ...989...81S}, where it was shown that $\sim 10^4$ events can put tight constraints on baryonic feedback models. 
All methods that rely directly on the Macquart relation ultimately measure the impact of feedback in this way, where the main contribution to the signal comes from uncertainty, but spatial information is largely integrated out.

\subsection{Locating the missing baryons}
The total baryon content of the Universe can be predicted under the $\Lambda$CDM framework using Big Bang nucleosynthesis, and the cosmic microwave background (CMB), which yields baryons to consist 5\% of the total energy budget \citep{cyburt_BBN_2016, cooke_determination_2018, aghanim_Planck_Cosmology_VI_2020}. Of these baryons, observations in the late Universe allocate $\sim$ 20\% to galaxies, and $\sim$ 50\% to warm-hot intergalactic medium (WHIM), leaving $\sim$30\% baryons unaccounted \citep{fukugita_cosmic_2004, shull_census_2012}. This discrepancy is called the `missing baryon' problem.

It is generally accepted that most of the missing baryons reside in a diffuse form in the IGM outside the dark matter haloes, which means detections are difficult due to low density contrast ($\delta \sim 10$), and temperatures of $10^{5 - 7}$K \citep{bregman_search_2007, mcquinn_evolution_2016}. Various methods searching for baryons in the IGM exist, such as X-ray emission and absorption \citep{bregman_searches_2009, eckert_warm_2015, nicastro_observations_2018, tanimura_stacked_2020}, the thermal Sunyaev-Zel’dovich (tSZ) effect in the filaments \citep{bonjean_gas_2018, degraff_probing_2019, tanimura_density_2020}, Lyman-$\alpha$ forest \citep{penton_local_2004}, and OVII absorption line \citep{nicastro_observations_2018}, which claim to resolve the issue by varying degrees of uncertainties. 

FRBs provide an independent probe for the baryon census. The characteristic dispersion in FRBs arises from the interaction of radio waves with intergalactic plasma and therefore traces ionised baryons. The IGM/cosmic contribution to the total DM encodes the most cosmological information. It is also the dominant factor compared to contributions from the Milky Way and the FRB host, which scale with redshift as $(1+z)^{-1}$.

With localised FRB samples at high redshifts, the DM-$z$ or Macquart relation \citep{macquart_census_2020} can directly probe the baryons in the intergalactic distances, where studies have already shown to provide constraints consistent with CMB \citep[see e.g.,][]{mcquinn_missing_2014, walters_future_2018, walters_probing_2019, li_cosmology_2019, macquart_census_2020, dai_reconstruction_2021, wang_determination_2023, 2025NatAs.tmp..131C, 2025arXiv250717742R}. These studies quantify the missing baryons in terms of the fractional baryon density $\Omega_\mathrm{b}$, where, with growing numbers of localised FRBs, the uncertainties have reduced. However, the question remains whether we can quantify the contribution in terms of the density contrast $\delta$ characterising the cosmic structures into voids, filaments, and haloes \citep[see e.g.,][for application with $N$-body simulation]{jaroszynski_fast_2019, zhang_intergalactic_2021, walker_dispersion_2024}. 

Generalising this process for any arbitrary number of partitions of the density contrast would lead to a linear equation of the form
\begin{equation}
\label{eq:linear_inversion}
    \mathbf{y} = \mathbf{M}\cdot\mathbf{x} + \mathbf{c} \;,
\end{equation} 
where $\mathbf{y} = \{\mathrm{DM}_1, \dots , \mathrm{DM}_n\}^\mathrm{T}$ is the observed dispersion measure, $\mathbf{M}$ is the mixing matrix, $\mathbf{x}$ is a binned function of the density contrast. Finally, in this model, $\mathbf{c}$ represents the noise as an additional additive term. For real observation, one has access to observed DM, and the mixing matrix can be constructed for a given fiducial cosmology and redshift bins. Inverting the linear equation yields the underlying density estimator.

However, in the general case, one has more DM values than the number of partitions in $\mathbf{x}$, which constitutes an over-determined system, rendering the inversion of \Cref{eq:linear_inversion} ill-posed. Moreover, another challenge with this approach is that the individual signals are noisy, as we lack comprehensive knowledge of the contributions from the FRB host galaxies and their CGM, which necessitates a statistical analysis. The key characteristics of such a sample are the host localisation and the correlations between them. An independent measurement of the redshift via host localisation is mandated to validate the stability of the framework with $N$-body simulations. The line-of-sight for two highly correlated FRBs \citep{reischke_cosmological_2023} would probe the same cosmic structure, which introduces further noise to the mixing matrix and, in turn, the inversion. Hence, a statistically large sample of FRBs ($\gtrsim$ 500) is required for a stable inversion.

The SKA survey aims to provide such a sample largely owing to the multi-beaming capabilities, where we expect to uncover a new region in the transient parameter space \citep[e.g., see Figure 3 in][]{macquart_fast_2015}. Additionally, with a sensitivity of 2 mJy in the 50-350 MHz band, SKA-LOW would detect FRB signals characterised by a coherent radiation of millisecond duration with a brightness temperature beyond the Compton limit \citep{kellerman_opaque_1969, tsang_inv_compton_limit_2007, fialkov_fast_2017}. This is critical for localising the host or for further warranting follow-up optical searches.

\subsection{Simulation-based Inference}
\begin{figure}
    \centering
    \includegraphics[width=0.65\linewidth]{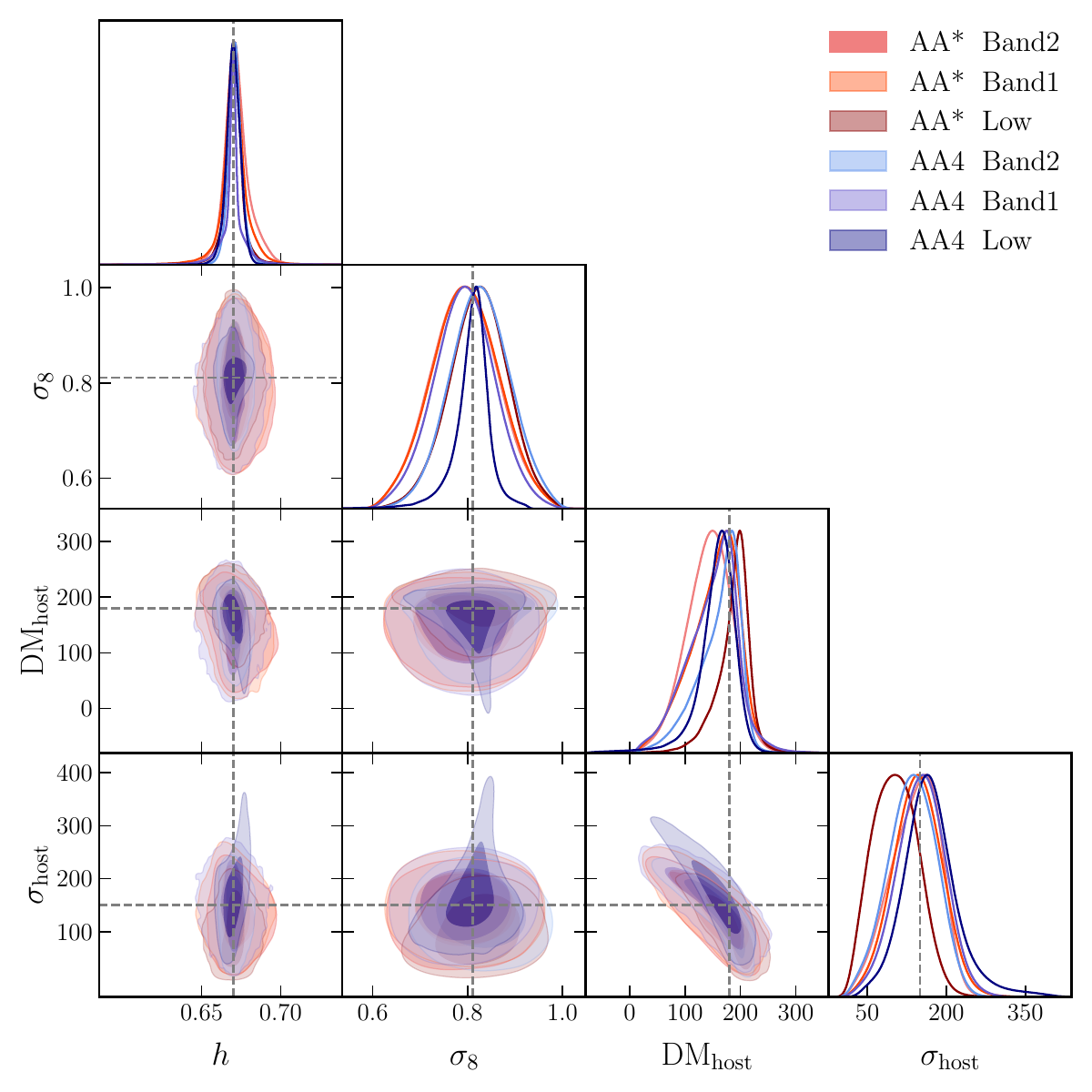}
    \caption{\small Marginalised contour plots for all the parameters in our inference process, with the shaded regions being the $1\sigma$ errors. Different colours indicate different SKA configurations. The grey lines represent the fiducial values of the mock catalogue. The results, except for AA4 low, are at $N_\mathrm{side} = 512$. The contours for AA4 low are from simulations with $N_\mathrm{side} = 1024$, and can therefore constrain the $\sigma_8$ parameter.}    
    \label{fig:contour_sbi}
    \vspace{.3cm}
\end{figure}
As discussed, previous analyses neglect the correlation between different sightlines. The main reason is that modelling this process is very cumbersome. While \citet{reischke_cosmological_2023} provided an estimate for the covariance between different sightlines, this captures the whole story only for a Gaussian likelihood. However, we know that the likelihood is more closely approximated by a log-normal distribution or a slightly different distribution \citep{jaroszynski_fast_2019,2025arXiv250707090K}. Implementing correlations in those distributions with analytical methods is not feasible. Therefore, one requires accurate forward simulations which can be generated quickly. This is then followed by simulation-based inference (SBI), an inference process that does not require an explicit likelihood model \citep{papamakarios_SNPE_2016, cranmer_SBI_2020}. It employs neural networks (NN) to learn the joint probability of the data-parameter pair by approximating the posterior $p(\boldsymbol{\theta}|\mathbf{d})$ with a conditional density estimator $q_\phi(\boldsymbol{\theta}|\mathbf{d})$, where $\phi$ are the trainable parameters. Based on the specific use case, the SBI algorithms and their respective loss functions differ\footnote{A comprehensive list of various algorithms and their implementations can be found at \href{https://github.com/sbi-dev/sbi}{https://github.com/sbi-dev/sbi}.}, which are broadly classified into two categories. The first kind is called the amortised method, where the final posterior can be utilised for many observations without retraining. In contrast, sequential algorithms are tailored to specific observations to improve simulation efficiency.

As the SBI framework requires forward-simulated data, we implement the prescription outlined in \citet{konar_constrain_2025}. For a given cosmological model, we create concentric spherical projections of the matter field in increasing redshift via \texttt{GLASS} \citep{tessore_glass_2025}, which accounts for the non-linear evolution via the log-normal realisation of the said matter field. For a localised FRB with a given sky position, the DM is the line-of-sight integral through the spherical shells, which constitutes the simulated data. However, as we currently lack observations from a real catalogue, we create a mock catalogue. The RA values are random samples from $\mathcal{U}[0, 2\pi)$ and the DEC values are chosen from $\arcsin(\,\mathcal{U}[-1,1)]\,)$ to maintain uniformity on a sphere. The rest-frame FRB host is modelled after a log-normal distribution of the form
\begin{equation}
\label{eq:lognormal_pdf}
   p\,(\mathrm{DM}_\mathrm{host,\,rf}; \mu, \sigma_\mathrm{host}) = \frac{1}{\mathrm{DM}_\mathrm{host,\,rf}\,\sigma_\mathrm{host}\,\sqrt{2 \pi}} \, \mathrm{exp}\,\left[-\,\frac{1}{2} \,\left(\frac{\ln\, (\mathrm{DM}_\mathrm{host,\,rf}) - \mu}{\sigma_\mathrm{host}}\right)^2 \right]\;,
\end{equation}
where the median ($\mathrm{exp}(\mu)$) and the scale ($\sigma_\mathrm{host}$) are the free parameters. Furthermore, the Hubble parameter ($h$), which describes the expansion rate, and $\sigma_8$, which quantifies matter clustering on scales of 8 $h^{-1}$ Mpc, are two additional cosmological parameters to be constrained in our model.

For different SKA configurations, we create mock catalogues based on the localised FRB numbers for one year. For our analysis, the prior ranges are $h=\mathcal{U}[0.4, 1]$, $\sigma_8 = \mathcal{U}[0.6, 1]$, $\mathrm{DM}_\mathrm{host} = \mathcal{U}[10, 1500]$ and $\sigma_\mathrm{host} = \mathcal{U}[10, 800]$. The SBI finally yields the corresponding posterior distributions. The results are shown as marginalised contour plots in \Cref{fig:contour_sbi}. The forward simulations are at $N_\mathrm{side} = 512$, corresponding to a multipole of $\ell = 3 \, N_\mathrm{side} - 1 \approx 1500$. However, note that the matter fields as a function of $\sigma_8$ are obtained via a spline and are only sensitive at $N_\mathrm{side} \geq 1024$ or smaller scales. This is highlighted by the final contour corresponding to the AA4 low configuration, for which we simulate at $N_\mathrm{side} = 1024$ and show that $\sigma_8$ can be constrained.
As the SKA approaches, this procedure will become the gold standard for reliably inferring cosmological or astrophysical parameters from FRBs.

\begin{figure}
    \centering
    \includegraphics[width=0.99\linewidth]{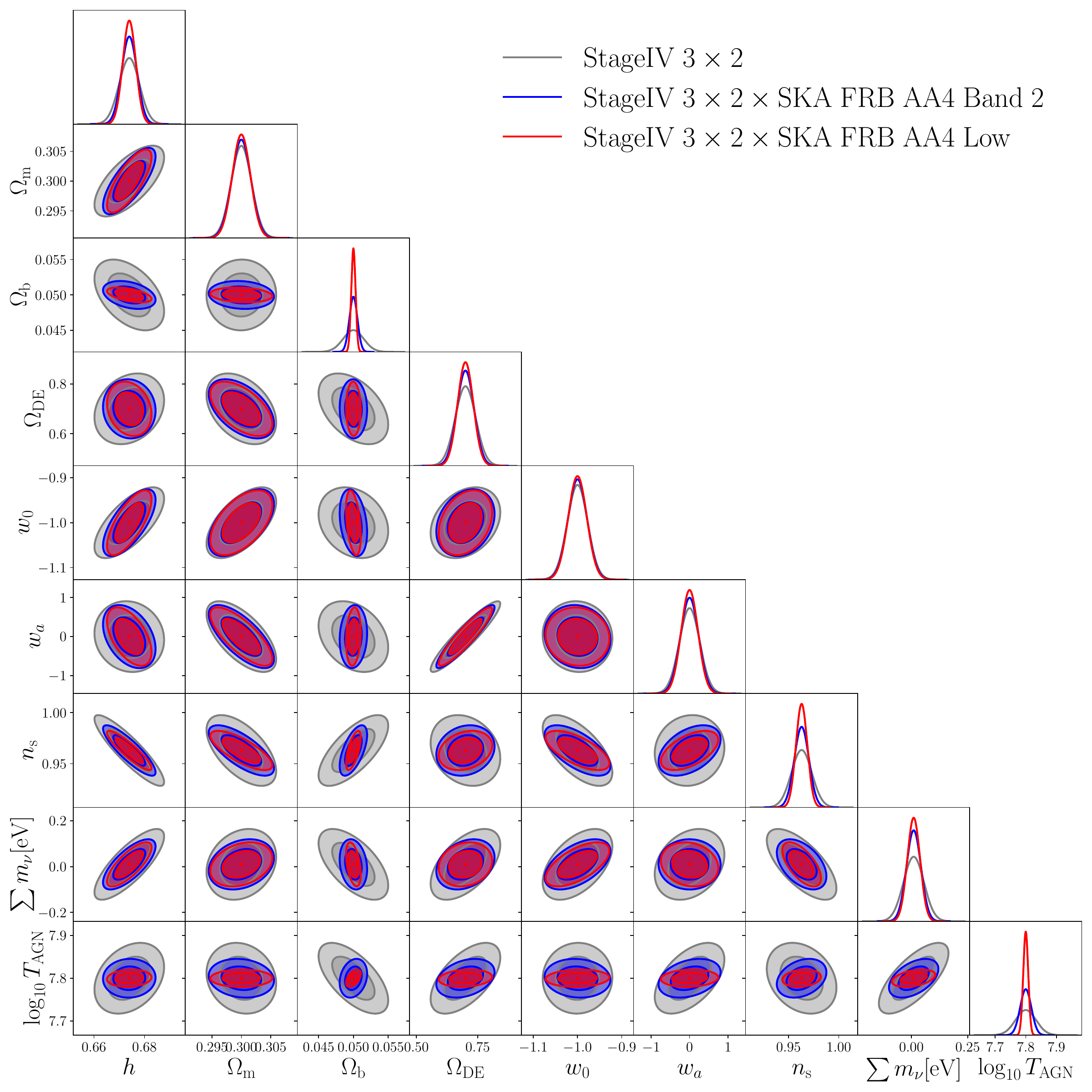}
    \caption{Marginal constraints on cosmological parameters. All contours show the one and 2$\sigma$ constraints. Grey contours denote a $3\times 2$ analysis with a Stage IV galaxy survey, such as Rubin-LSST or Euclid. The blue and red contours show the improvement achieved by adding 5 years of observations from AA4 in Band 2 and Low, respectively.}
    \label{fig:frbs_stage_4aa4}
\end{figure}

\section{Correlations in the dispersion measure}
\label{sec:two_point_feedback}
In the previous sections, we have examined the variance in the Macquart relation and what it reveals about the distribution of baryons. This, however, is an integral over the electron power spectrum and hence wipes out the spatial information. Here, we thus discuss how the spatial information of the baryon distribution traced by FRBs can be used in cosmological inference.

\subsection{Two-point function analysis in synergy with Stage IV galaxy surveys}
The simplest summary statistic of LSS observables is the angular two-point function or its harmonic-space equivalent, the angular power spectrum, $C(\ell)$. For any two projected tracers $A$ and $B$ the angular power spectrum \citep[assuming the Limber approximation][]{2008PhRvD..78l3506L} is
\begin{equation}
   C_{AB}(\ell)  = \int_0^{\chi_\mathrm{H}}\frac{\mathrm{d}\chi}{\chi^2}W_A(\chi) W_B(\chi) P_{ab}\left(\frac{\ell + 1/2}{\chi},z(\chi)\right).
\end{equation}
Here, $W_A$ are probe-specific weights, $P_{ab}$ is the 3D power spectrum of the two probes, and $\chi$ is the comoving distance. For FRBs, $W_A(\chi)$ takes the following form \citep[e.g.][]{2021PhRvD.103b3517R}:
\begin{equation}
     W_\mathrm{DM}(\chi) = \frac{3 H_0^2 \Omega_{\mathrm{b0}}}{8\pi G m_\mathrm{p}}\chi_\mathrm{e}[1+z(\chi)] f(z(\chi))\int_\chi^{\chi_\mathrm{H}}\mathrm{d}\chi'\, n_\mathrm{FRB}(\chi'),
\end{equation}
where $n_\mathrm{FRB}$ is the normalised distribution of observed FRBs. The SKA FRB detection counts and their redshift distribution require a number of ingredients: instrument properties, an intrinsic FRB population model, and a cosmological ionised-gas model. Specific details of this are outlined in the FRB chapter of the transient science working group. In the same chapter, the host detectability is discussed: while FRB follow-up can cover only a small fraction of detections. However, SKA and LSST will be highly complementary, and host visibility can be estimated assuming a Stage IV galaxy survey source density. The product of $n_\mathrm{FRB}$ and the detection probability provides the effective redshift distribution of FRBs. In principle, however, one can also auto-calibrate the redshift distribution using calibrated photometric samples \citep{2023arXiv230909766R}.
To further increase the synergy with Stage IV galaxy survey cosmological photometric surveys, we consider a standard $3\times 2$pt analysis and include cosmic shear and photometric galaxy clustering with the following weight functions 
\begin{equation}
\begin{split}
        W_{\kappa_i}(\chi) = & \,\frac{3\Omega_\mathrm{m0}}{2\chi^2_\mathrm{H}}\frac{\chi}{a(\chi)}\int_\chi^{\chi_\mathrm{H}}\mathrm{d}\chi'n^{s}_i(\chi')\frac{\chi-\chi'}{\chi'}\,.\\
        W_{\mathrm{g}_i}  = &\, n^{l}_i(\chi)b_i(z(\chi)),
\end{split}
\end{equation}
respectively. Here, $n^{l}_i(\chi)$ and $n^{s}_i(\chi)$ are the lens and source samples in tomographic bin $i$. For the bias, $b_i$, we assume a single free parameter in each of the ten tomographic bins. All settings are taken from the optimistic setting presented in \citet{2020A&A...642A.191E}.

\begin{figure}
    \centering
    \includegraphics[width=0.99\linewidth]{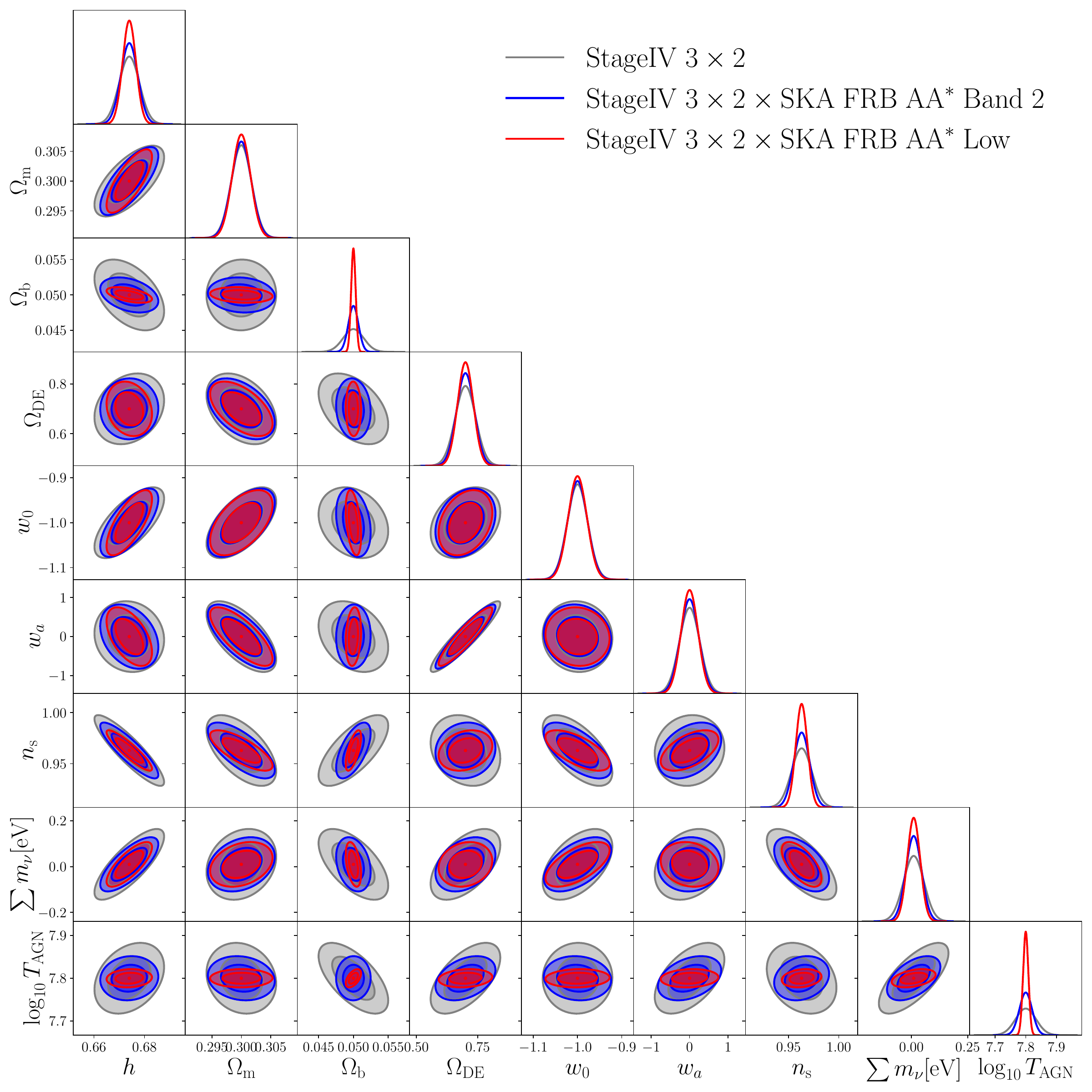}
    \caption{Marginal constraints on cosmological parameters. Same colour scheme as \Cref{fig:frbs_stage_4aa4} but for AA$^\star$ instead of AA$4$.}
    \label{fig:frbs_stage_4aastar}
\end{figure}

Any observed spectrum, $\hat{C}_{AB}(\ell)$, contains a white-noise component from the discreteness of the tracer:
\begin{equation}
    \hat{C}_{AB}(\ell) = {C}_{AB}(\ell) + N_{A}\delta^\mathrm{K}_{AB},
\end{equation}
with Kronecker delta $\delta^\mathrm{K}_{AB}$ and noise level $N_{A}$. The noise levels for cosmic shear and FRBs are
\begin{equation}
\label{eq:noiselevels}
\renewcommand{\arraystretch}{1.2}
    N_{A} = 
    \left\{\begin{array}{cl}
        \frac{\sigma^2_{\epsilon,i}}{2\bar{n}^{s}_i} & \quad \mathrm{if} \quad A = \kappa_i\\
        \frac{\sigma^2_\mathrm{host} + \sigma^2_\mathrm{DM}}{\bar{n}^\mathrm{FRB}(1+\bar{z})} & \quad \mathrm{if} \quad A = \mathrm{DM}\\
        \frac{1}{\bar{n}^{l}_i} & \quad\mathrm{if} \quad A = \mathrm{g}_i
    \end{array}\right.
\end{equation}
where $\sigma_{\epsilon,i} = 0.26$ is the total ellipticity dispersion of the source sample in tomographic bin $i$ with mean number density $\bar{n}^{s}_i$ and $\bar{n}^\mathrm{FRB}$ and $\bar{z}$ are the mean number density and mean redshift of the FRB sample. $\frac{1}{\bar{n}^{l}_i}$ is the number density of lenses used for the clustering measurements. The
\Cref{eq:noiselevels} highlights a key advantage of FRBs over cosmic shear: galaxy intrinsic ellipticity greatly exceeds the cosmological signal, hence no term $\sigma^2_{\kappa_i}$. For FRBs, the intrinsic noise is lower than the signal from $z\gtrsim 0.5$. The variance of the DM is given by
\begin{equation}
    \sigma^2_\mathrm{DM} = \int \frac{\ell \mathrm{d}\ell}{2\pi} C_{\mathrm{DMDM}}(\ell). 
\end{equation}
Since the cosmological contribution of the DM carries direct information on the electron density, measuring different summary statistics of the DM opens up a new look at the baryonic Universe \citep[e.g.][]{mcquinn_missing_2014,masui_dispersion_2015,2021ApJ...922...42R,bhattacharya_fast_2021,2021MNRAS.502.2615T,2023arXiv230909766R,2025NatAs.tmp..131C, 2025ApJ...989...81S,2025arXiv250604186H,2025ApJ...983...46M,2025A&A...698A.163H}. 

We forecast the constraints using Fisher matrices. To this end, we stack bundle (in bin $\ell_\alpha$) observed spectra into a data vector:
\begin{equation}
    \boldsymbol{C}^\mathrm{T}(\ell_\alpha) = \left(\{\hat{C}_{AB}(\ell_\alpha)|\;\mathrm{all\;unique\;combinations\;}AB \;\mathrm{with\;} A,B\in(\kappa_i, \mathrm{DM}, \mathrm{g}_i) \}\right).
\end{equation}
Here, we distinguish between two analyses: a classical $3\times 2$ analysis of a Stage IV galaxy survey that includes cosmic shear, galaxy clustering, and galaxy-galaxy lensing. Secondly, a $3\times 2$ matrix that includes the DM observed across different SKA settings (i.e., different bands or configurations). This adds $3$ additional spectra, namely the correlations DMDM, $g_i$DM, and $\kappa_i$DM, resulting in a $6\times 2$ analysis. We then assume a Gaussian covariance $\boldsymbol{\mathrm{C}}$, so that:
\begin{equation}
    \boldsymbol{\mathrm{C}}_{ABCD} = \mathrm{cov}(\hat{C}_{AB}(\ell_\alpha),\hat{C}_{CD}(\ell_\alpha)) = \frac{4\pi\left[\hat{C}_{AC}(\ell_\alpha)\hat{C}_{BD}(\ell_\alpha) + \hat{C}_{AD}(\ell_\alpha)\hat{C}_{BC}(\ell_\alpha)\right]}{\mathrm{max}(\mathcal{A}_{AB}, \mathcal{A}_{CD})(2\ell_\alpha + 1)\Delta \ell_\alpha}\;,
\end{equation}
where we have already used that different $\ ell$-modes are uncoupled in the Gaussian limit. $\mathcal{A}_{AB}$ is the survey area over which the combination $A,B$ is measured $\ell_\alpha$ is the centre of the multipole bin and $\Delta\ell_\alpha$ its width.
We then calculate the components of the Fisher matrix as:
\begin{equation}
    F_{ij} = \sum_\alpha\partial_i \boldsymbol{C}^\mathrm{T}(\ell_\alpha)\boldsymbol{\mathrm{C}}^{-1}(\ell_\alpha)\partial_j \boldsymbol{C}(\ell_\alpha),
\end{equation}
where $\partial_i$ indicates the partial derivative with respect to the $i$-th cosmological parameters. We assume a survey area of $15\,000\,\mathrm{deg}^2$. While SKA can detect FRBs over a larger area, we cut the Milky Way to avoid the uncertain modelling of $\mathrm{DM}_\mathrm{MW}$ in the disk. The cross-correlations will also be measured over the same area. Multipoles are collected up to $5000$ for cosmic shear and the DM. Galaxy clustering is limited to $500$ by the linear bias assumption. 

We display the results of this exercise in \Cref{fig:frbs_stage_4aa4,fig:frbs_stage_4aastar} for the AA4 and AA$^*$ configuration, respectively. All contours are shown as one and two $\sigma$ intervals. The grey contours depict a standard 3$\times 2$ analysis as being carried out by Stage IV galaxy surveys (corresponding to Euclid DR3 or Rubin-LSST Y10). In blue and red, we show how the constraints change when we add FRBs detected from SKA in Band 2 and Low, respectively. \review{For other studies considering the statistical properties of FRBs in conjunction with LSS tracers or general studies beyond vanilla $3\times 2$pt, we refer to \citep{2023arXiv230909766R,2025arXiv250919514L,2026ApJ...998..109S,2026arXiv260212174W}.}
For SKA, we use 5 years' worth of commensal observations and adopt the most optimistic FRB count from the synthetic redshift distribution \citep[see Section 3 in][]{Caleb02.2026.SKA}.

\begin{figure*}
    \centering
    \includegraphics[trim={0cm 0cm 1cm 0cm}, clip, width = .49\textwidth]{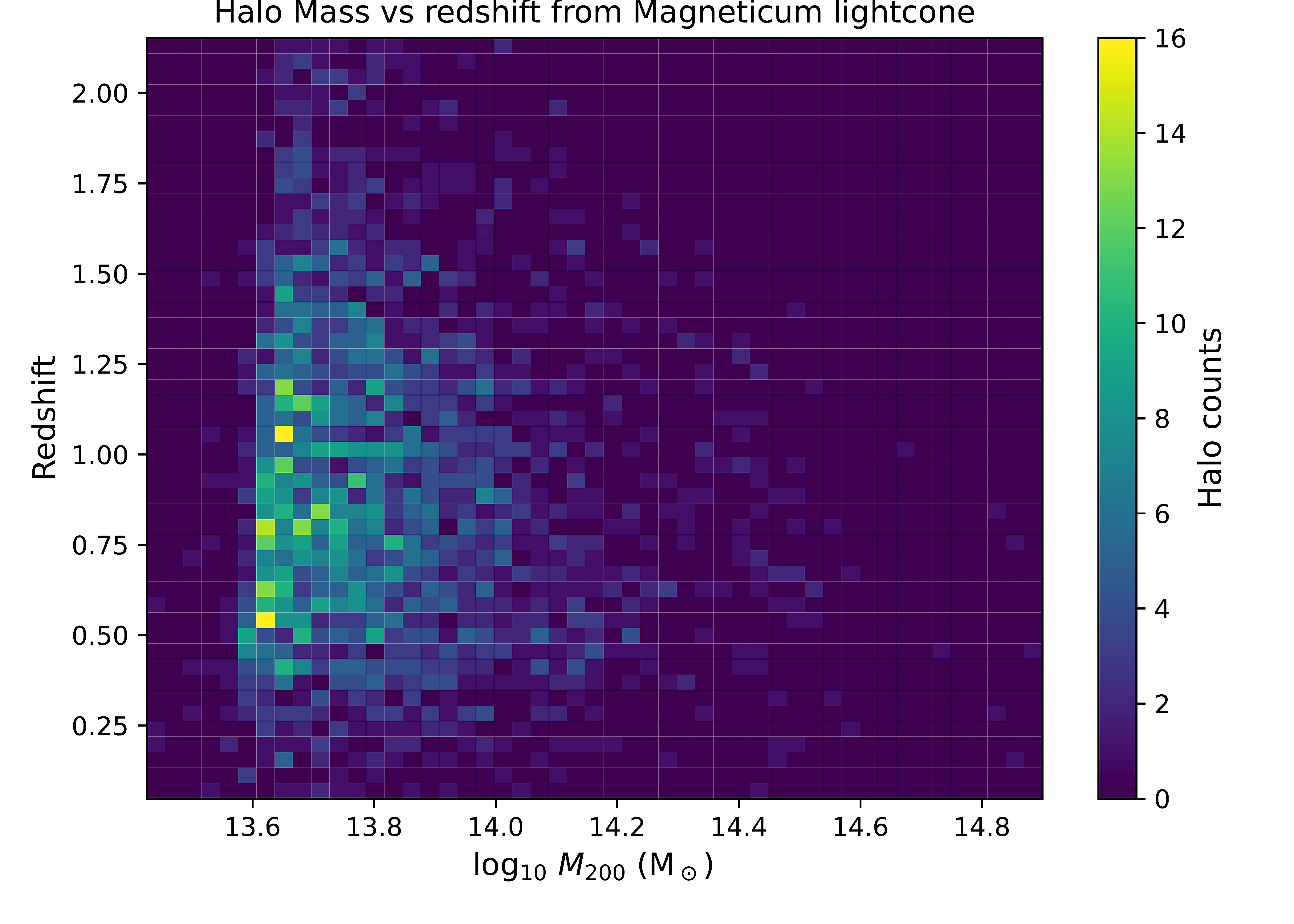}
    \includegraphics[trim={1cm -.2cm 0cm .2cm}, clip,width = .49\textwidth]{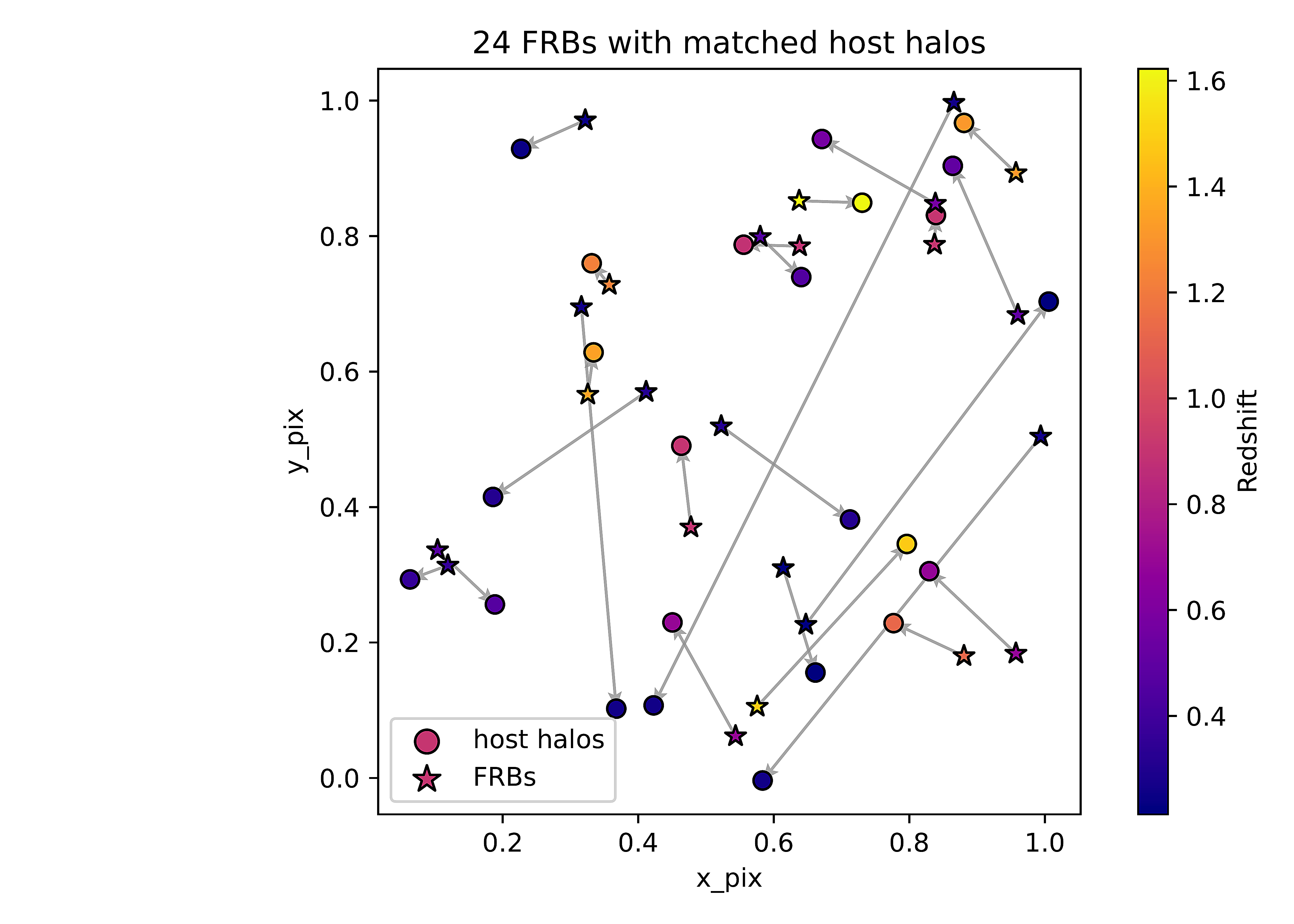}
    \caption{{Left: Mass and redshift distribution of halos from Magneticum lightcone (25 square-degrees). Right: FRBs (from AA* predictions) matched to simulated halos.}}
    \label{fig:M200_z_magneticum}
\end{figure*}

In general, we can make the following observations: SKA is particularly well-suited to improving constraints on the parameters controlling the overall amplitude of the DM auto-correlation, i.e., $h$ and $\Omega_\mathrm{b}$. On the other hand, $\Omega_\mathrm{m}$, which controls the amplitude of the lensing signal, is not improved. This also shows that the overall signal-to-noise ratio of a Stage IV galaxy survey is higher than that of DM correlations from SKA. Apart from these amplitude parameters, one can see that FRBs shine at constraining baryonic feedback via $\log_{10}T_\mathrm{AGN}$. This also improves the sensitivity to other parameters, changing the shape of the small-scale power spectrum ($n_\mathrm{s}$ and $\sum m_\nu$). Overall, we find that SKA can improve these constraints by a factor of $2-5$ depending on the parameter. Providing prior information on $\Omega_\mathrm{b}$ or $h$ from CMB measurements will further enhance the gain from SKA due to the degeneracy breaking between the baryon density, the Hubble constant and the amount of feedback. Generally, we observe that numbers outperform redshift coverage. The reason for this is twofold: first, a larger number of FRBs reduces the shot-noise contribution. While a larger redshift means a larger contribution to the DM from the LSS, reducing the host noise, the field variance in the shot noise is independent of this effect. Secondly, the Low configuration of SKA upweights the relative contribution from small redshifts and smaller scales, i.e., precisely the regime where baryonic feedback is extremely important.

\subsection{Cross-matching halos }
  Most of the baryons reside in the CGM, a multi-phase gaseous reservoir that surrounds the central disk and extends at least to the galaxy's virial radius. For halo masses above $ 10^{11} M_\odot$, the ionised CGM contains the bulk of the baryonic mass of the galaxy. However, the total mass, density, and temperature distribution of the CGM remain poorly constrained. 
Both simulations \citep{2023MNRAS.524.5391A,2025ApJ...980...61M} and observational \citep{2023ApJ...951..125D,2024A&A...690A.268Z} studies consistently show that the CGM can be redistributed much beyond the virial radius of galaxies during robust feedback processes, modifying its spatial distribution in the process. 
A possibility with SKA is to directly identify halos and attempt to measure their profiles and mass fractions. 
\review{The approach done by FLIMFLAM \citep{2024ApJ...973..151K} could be adapted and extended to the SKA context, that is {variance reduction by reconstruction} could exploit the dense spectroscopic and HI galaxy catalogues that the SKA will provide: by performing Bayesian matter-density reconstructions along each FRB sightline, the cosmic variance that currently dominates the scatter in the Macquart relation would be substantially suppressed, sharpening the separation of the diffuse IGM contribution from the localised CGM signal. As illustrated on the right side of \Cref{fig:M200_z_magneticum}, mock FRB localisations in a simulation field can already be reliably matched to their host halos across a broad redshift range $z \sim 0.3$--$1.6$, demonstrating that the host identification needed for this reconstruction step is feasible over cosmologically interesting volumes. }

\review{
On the other hand, one could stack profiles directly, which would become particularly powerful with the large halo populations accessible to the SKA. The left side of \Cref{fig:M200_z_magneticum} shows that the Magneticum lightcone contains a rich population of halos concentrated at $\log_{10}(M_{200}/\mathrm{M}_\odot) \sim 13.5$\textendash$14.0$ and $z \sim 0.4$\textendash$1.2$, with counts per bin reaching $\gtrsim 10$. By modelling the radial electron-density profile of each identified halo and stacking hundreds of intersections binned by $M_{200}$ and redshift, one would obtain statistically robust CGM gas-fraction and density-profile measurements extending well below the regime accessible to X-ray or Sunyaev--Zel'dovich observations, and out to and beyond the virial radius where feedback-driven baryon redistribution is expected to be most significant. The combination of both techniques, using reconstruction to remove the large-scale IGM variance while simultaneously forward-modelling individual halo crossings, would allow the SKA to deliver joint constraints on $f_{\mathrm{igm}}$ and $f_{\mathrm{gas}}(M_{200}, z)$ with a precision that is currently unattainable.}

\section{Cool gas abundance and morphology in the CGM}
\label{sec:CGMscattering}

Quasar absorption measurements have established the ubiquitous presence of a cool ($T \sim 10^4\,
{\rm K}$) gas phase in the CGM of many galaxies across a range of masses and redshifts. This cool gas apparently has a low volume filling fraction, but a high covering fraction out to large radii, potentially beyond the virial radius \citep{Tumlinson2017CGMReview, Rudie2019}. The presence of this cool phase, especially at large radius and large halo mass, has challenged standard pictures of cool accretion, whereby the cool phase should be suppressed above some halo mass \citep{Tumlinson2017CGMReview, FaucherGiguere2023Review}. Physical models have been proposed to explain the existence of this gas; however, progress has been challenging due to the small physical scales involved. The high covering fraction and low volume filling fraction imply a characteristic size for the cool gas absorbers of roughly parsec scale \citep{McCourt2018}. This is well below the resolution achievable even in state-of-the-art simulations, i.e., the subgrid physics.

FRBs may be scattered by cool gas in the halos of intervening galaxies. FRB scattering is a complementary probe of gas in the CGM to DM measurements. While DM is an integrated measure, sensitive to the total electron column along the line of sight, scattering results from multi-path propagation due to transverse inhomogeneities in the electron density. In particular, scattering may be sensitive to inhomogeneities down to $\sim {\rm au}$ scales in the CGM of intervening galaxies \citep{VP2019}. Scattering is sensitive to not only the abundance of cool gas and the amplitude of density fluctuations in the gas \citep{Ocker2025CGMScattering}, but also the morphology of the gas \citep{Jow2024CGMScattering}. Thus, FRB scattering is a critical probe of cool gas, shedding light on the physics of the CGM on small scales. 

To date, no scattering or plasma lensing due to gas in the CGM has been confirmed. The SKA-Low provides an advancement in our ability to detect CGM scattering for two reasons: 1. FRBs at high redshift are almost guaranteed to intersect a halo within the virial radius, 2. as the strength of scattering scales as roughly $\sim \nu^{-4}$, the low frequency coverage of SKA-Low will result in a significant enhancement in scattering effects. To assess the prospects for detecting CGM scattering with SKA-Low, we will use the formalism for scattering by a turbulent medium employed by \citep{Ocker2025CGMScattering}. The scattering timescale due to a turbulent medium is given by
\begin{equation}
    \tau = 48\,{\rm ns}\,\frac{G_{\rm scatt}}{(1 + z_l)^3}\,\left(\frac{\tilde{F}_l}{({\rm pc^2 \, km)^{-1/3}}}\right) \left( \frac{\rm DM_l}{\rm pc \,cm^{-3}} \right)^2 \left( \frac{\nu}{\rm GHz} \right)^{-4},
    \label{eq:tauFl}
\end{equation}
where $z_l$ is the redshift of the scatterer, $\nu$ is the observing frequency, ${\rm DM}_l$ is the excess DM of the scatterer, and $G_{\rm scatt}$ is the geometric leverage parameter, $G_{\rm scatt.}(z_l,z_s) = D_{ls}D_l / 2 D_{s} L $, where $L$ is the thickness of the scattering material. The parameter, $\tilde{F}_l$, is a parameter characterizing the magnitude of the density fluctuations of the turbulent medium, and, for a Kolmogorov spectrum of density fluctuations, is given by $\tilde{F}_l = \zeta \epsilon^2 (l_o^2 l_i)^{-1/3}$, where $l_o$ is the outer-scale of the fluctuations, $l_i$ is the inner-scale, and $\zeta = \langle n_e^2 \rangle / \langle n_e \rangle^2$ and $\epsilon = \langle \delta n_e^2 \rangle / \langle n_e \rangle^2$. The value of this parameter for the cool absorbed in the CGM is poorly constrained (especially since it depends on the unknown inner scale of the turbulence), but it can be estimated under certain assumptions from quasar absorption measurements of the velocity fluctuations of the cool gas phase. In this way, \citet{Ocker2025CGMScattering} arrive at a benchmark value of $\tilde{F}_l \sim 0.5\times 10^{-3}\,({\rm pc^2\,km})^{-1/3}$.

\begin{figure}
    \centering
    \includegraphics[width=0.8\linewidth]{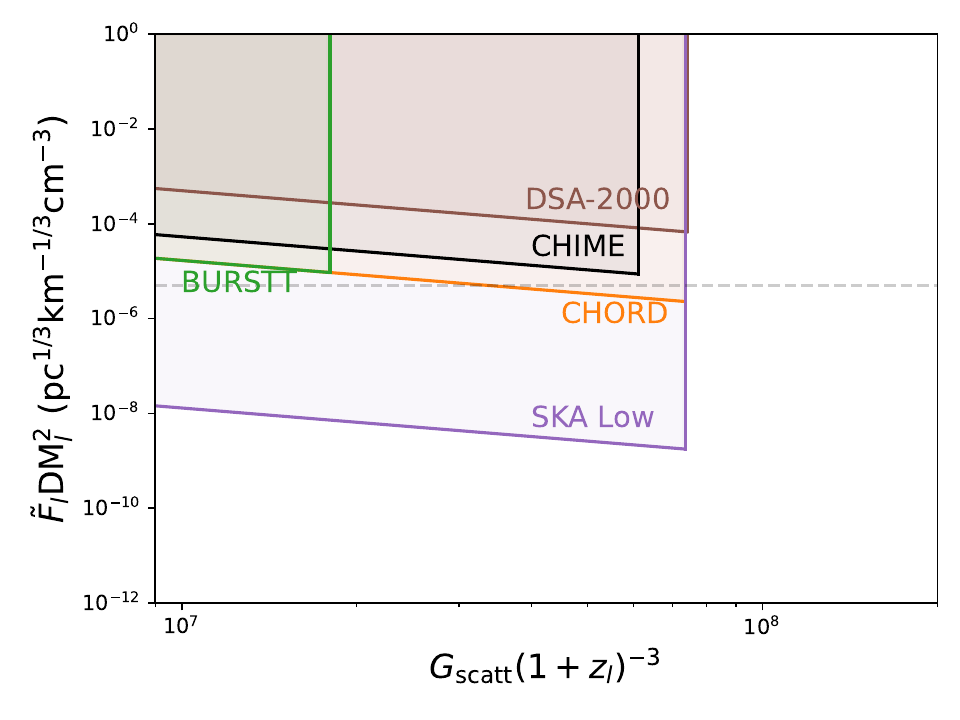}
    \caption{CGM scattering parameters, $\tilde{F}_l {\rm DM}_l^2$ and $G_{\rm scatt} (1 + z_l)^{-3}$, that will produce a $\tau > 1\,{\rm ms}$ scattering tail for BURSTT, CHIME, CHORD, DSA-2000, and SKA-Low. The grey, dashed line represents the fiducial value of the fluctuations expected for cool gas in the CGM, $\tilde{F}_l = 0.5 \times 10^{-3}\,(\rm pc^2\,km)^{-1/3}$ and ${\rm DM}_l = 0.1\,{\rm pc \,cm^{-3}}$. The geometric leverage $G_{\rm scatt}$ is computed assuming a size of $L = 10\,{\rm pc}$ for the cool-gas scatterer.}
    \label{fig:forecast}
\end{figure}

\Cref{fig:forecast} shows the sensitivity of different FRB experiments to scattering parametrised in this way. The lines correspond to $\tau = 1\,{\rm ms}$; i.e. within the shaded regions, the scattering from the CGM exceeds $1\,{\rm ms}$ for those values of $\tilde{F}_l \,{\rm DM}_l^2$ and $G_{\rm scatt} (1+z_l)^{-3}$. The different curves reach lower values of $\tilde{F}_l \,{\rm DM}_l^2$ due to the lower frequency coverage. 
The maximum value of $G_{\rm scatt} (1 + z_l)^{-3}$ is computed out to the maximum redshift in the simulation ($z=5$). 
Taking this as the maximum source redshift, we find the $z_l$ that maximises $G_{\rm scatt} (1 + z_l)^{-3}$, and take this as the right-most limit of the sensitivity curves. The equivalent maximum redshift for the other FRB experiments is estimated from their projected sensitivities. Note that the primary advantage of SKA-Low is in the frequency coverage. While higher redshift FRBs will, on average, have a higher geometric leverage for scattering, they will also have higher frequencies in the rest-frame of the scatterer (as represented by the $(1+z_l)^{-3}$ term in Eq.~\ref{eq:tauFl}). The primary advantage of higher redshifts is not in the coverage of the scattering parameter space, but in the fact that high redshift sources are almost guaranteed to intersect a galactic halo within the virial radius. The primary takeaway from \Cref{fig:forecast} is that while CGM scattering may be just beyond the detection ability of current experiments, such as CHIME, it will be well within the range of SKA-Low, for both AA$^*$ and AA4 configurations. The primary challenge for detection, therefore, will not be sensitivity but rather disentangling the contribution of the CGM to the observed scattering from that of the Milky Way ISM and the host environment. 

\section{Epochs of Reionisation}
\label{sec:eor}
Our Universe has undergone two periods of ``reionisation" during which either hydrogen or helium is photoionised. Hydrogen reionisation occurred at $6 \lesssim z \lesssim 25$, and the ionisation history during this period sheds light on the earliest photoionising astrophysical sources. Measuring the redshifted neutron hydrogen signal during this transition is a major focus of radio astronomy at $\lesssim$200 MHz, and there is an entire SKA Science Working Group dedicated to these efforts. 

\subsection{Hydrogen reionisation}

FRBs, which trace ionised matter, are complementary probes to 21-cm experiments that measure neutral hydrogen. Therefore, FRBs have high scientific potential, but it has not yet been demonstrated observationally that FRB central engines existed too early in the Universe. Therefore, predicting the number of FRBs during the hydrogen EoR is highly speculative, and the simulation software used to predict the FRB population discovered across the six array configurations and bands ended at a redshift of 5 (see \citet{Caleb02.2026.SKA} for details). Nevertheless, in the absence of any better guide, we use the statistics at $z<5$ to extrapolate to higher redshifts.
The predicted number of FRBs detected per accumulated year of observation at $z>6$ for the SKA-low and SKA-mid (bands 1 and 2) AA4 array is $\sim$1 and $\sim10$, respectively. We caution that these estimates assume the ensemble-average FRB-redshift distribution, and, as shown in \citep{Caleb02.2026.SKA}, there is significant variation in any given simulation realisation, particularly for SKA-low, where the input parameters are poorly constrained. For example, no redshift dependence for scattering was included, and some modelling suggests that 20\% of FRBs at $z>5$ will be significantly scattered ($>$5~ms) even at 1.4 GHz \citep{Ocker+22}. 

EoR science with FRBs falls into two broad classes: statistical inference from a large number of FRBs that does not require redshift information, or deep follow-up of a smaller number to obtain the redshift and line of sight properties.
The former approach, in which, for example, the maximum observed DM is measured \citep[e.g.][]{2021MNRAS.502.5134B}, is not feasible with either AA* or AA4. Instead, maximising FRB-EoR science with the SKA will require deep, dedicated observational follow-up, likely with the most powerful optical or IR instruments. 
\citet{2021MNRAS.502.5134B} propose that the ionisation fraction in four redshift bins can be constrained with roughly 40 FRBs detected between $z=6- 9$. This is within the reach of the SKA, assuming the redshifts are measurable, but on a timescale of roughly a decade. More precise constraints on the ionisation history, or constraints on the CMB optical depth, could be made with of order 100 FRBs \citep{2022ApJ...933...57H,2024A&A...689A.340M}, but this will likely not be possible with the currently planned SKA configurations. 

\subsection{Helium reionisation}

Our Universe also underwent a period of helium reionisation, during which singly ionised helium (He\textsc{II}) was photoionised to He\textsc{III} at $z = 3-4$. This transition was primarily driven by hard UV and soft X-ray photons from quasars and other active galactic nuclei, and its timing and duration encode the thermal and ionisation history of the intergalactic medium. While helium reionisation cannot be probed directly via 21 cm emission, FRBs provide a complementary tool. During helium reionisation, the number of free electrons in the IGM increases because He\textsc{II} is ionised to He\textsc{III}, with each helium atom contributing an additional free electron. This increase produces a measurable enhancement in the DM for FRBs located at redshifts just above the He\textsc{II} reionisation epoch.

The signature of helium reionisation in FRB data is thus twofold. First, there is a redshift-dependent increase in the average IGM contribution to DM: FRBs beyond $z\sim3$ traverse regions where helium has been doubly ionised, leading to systematically higher electron column densities than lines of sight at slightly lower redshift. The effect is subtle, with a roughly 5-10\% increase in IGM DM, but it becomes detectable with sufficiently large FRB samples. Second, helium reionisation is believed to be patchy, with ionised bubbles forming around individual quasars before percolating across the IGM. This patchiness introduces line-of-sight variations in DM, as FRBs that happen to pass through He\textsc{III}-rich regions will have slightly higher DMs than those travelling through regions still dominated by He\textsc{II}. Measuring the scatter in DM at a given redshift thus allows statistical characterisation of the patchiness and morphology of helium reionisation.

The SKA will be well-suited to studying these signatures. Forecasts indicate that SKA-mid will detect up to $10^5$ FRBs over a few years, providing the large, high-redshift sample necessary to robustly measure the subtle DM increase and its scatter due to helium reionisation. The SKA is expected to detect the He\textsc{II} reionisation signal with high significance, achieving signal-to-noise ratios of $30-50\sigma$ for the amplitude of the DM step associated with He\textsc{II}.

\section{Conclusions}
\label{sec:conclusions}

FRBs enable a completely new way to probe the baryon distribution in the Universe, largely without any additional assumption about their temperature or pressure, making them a unique probe. The SKA is expected to substantially enhance our knowledge of the baryon distribution, even in very low-density environments inaccessible to other observations, by its imprint on FRB DMs. The width and amplitude of the Macquart relation will enable accurate measurements of the overall baryon density and its distribution, as outlined in \Cref{sec:macquart_relation_feedback}.

A connection to Stage IV galaxy surveys like Euclid and LSST will be highly beneficial. Those surveys will be limited by uncertainty in feedback models, which are essential at small scales where most of the expected signal originates. The SKA will be able to detect the 2-point correlation of the DMs with high significance, thereby adding to the other cosmological 2-point correlation functions (based on shear and galaxy clustering measurements). As shown in \Cref{sec:two_point_feedback}, the combination of these probes dramatically increases the cosmological information accessible by calibrating feedback and therefore the amplitude of the matter power spectrum at small scales.

The scatter-broadening of FRBs due to interactions with dense, cool gas in the CGM is a promising avenue to probe the state of the gas. The SKA's wide frequency coverage and high sensitivity for detecting high-z events (for which scattering is increasingly likely) present unique opportunities.

Lastly, studies of the epoch of reionisation using FRBs are promising but remain highly speculative at this point due to large uncertainties in the FRB population at high redshifts, as explained in \Cref{sec:eor}. Helium reionisation is well within the reach of the SKA and will be detected with high significance.

\textbf{Dedicated to the memory of J.P. Macquart, who wrote the original SKA FRB Science Book chapter “Fast Transients at Cosmological Distances with the SKA”.}

\section{Acknowledgments}

AS thanks Yik Ki (Jackie) Ma for several useful discussions. AS acknowledges support from the Australian Research Council's Discovery Early Career Researcher Award (DECRA, project DE250100003) and the Australia-Germany Joint Research Cooperation Scheme of Universities Australia (UA--DAAD, 2025--2026). LGS is a Lise Meitner Research Group leader and acknowledges funding from the Max Planck Society. JJ-S acknowledges support through the Australian Research Council Discovery Project DP220102305.
 D.M. acknowledges support from the French government under the France 2030 investment plan, as part of the Initiative d'Excellence d'Aix-Marseille Universit\'e -- A*MIDEX (AMX-23-CEI-088). JJ-S acknowledges support through the Australian Research Council Discovery Project DP220102305.

\bibliographystyle{abbrvnat-maxbibnames4}
\bibliography{chapter} 

\end{document}